\documentclass[proof]{WileyASNA-v1}

\usepackage[utf8]{inputenc}

\articletype{Article Type}%

\received{1 October 2020}
\revised{}
\accepted{}

\begin{document}

\title{A numerical approach for radiative cooling in relativistic outflows}

\author[1]{Jes\'us M. Rueda-Becerril*}

\authormark{\textsc{J. M. Rueda-Becerril}}

\address[1]{\orgdiv{Department of Physics \& Astronomy}, \orgname{Purdue University}, \orgaddress{West Lafayette, \state{IN}, \country{USA}}}

\corres{*Jes\'us M. Rueda-Becerril, Department of Physics, Purdue University, 525 Northwestern Avenue, West Lafayette, IN, 47907, USA. \email{jruedabe@purdue.edu}}

\abstract{In high energy astrophysics scenarios such as blazars, GRBs or PWNe, it is highly probable that ultra-relativistic particles interact with photons in their environment through scattering. As long as the energy of the particle is greater than the energy of the interacting photon, the (classical) scattering is known to be in the Thomson regime. Otherwise, quantum effects will affect the scattering cross section, and we enter into the so-called Klein-Nishina regime. It is well known that radiative cooling in the Thomson regime is very efficient, leading to soft high-energy spectra. However, observations have shown that, in many cases, the high energy spectrum of some objects is rather hard. This has led to think that maybe particles are not being cooled down efficiently. Asymptotic approximations of the Klein-Nishina regime have been formulated in the last decades in order to account for these corrections in the distribution of particles responsible for the observed spectrum of high energy sources. In this work we presenta a numerical approach of the Klein-Nishina corrections to the radiative cooling. It has been developed to simulate the evolution of a distribution of particles interacting with photons in their surroundings via inverse Compton scattering.
}

\keywords{methods: numerical, radiation mechanisms: non-thermal, acceleration of particles}

\maketitle

\footnotetext{\textbf{Abbreviations:} ANA, anti-nuclear antibodies; APC, antigen-presenting cells; IRF, interferon regulatory factor}

\section{Introduction}\label{sec:intro}

Gamma-ray bursts (GRBs) are among the most energetic objects in the Universe observed to date. According to their duration, these objects are mainly classified into: long ($\gtrsim 2$~s), and short ($\lesssim 2$~s). The former are found in star-formation regions, produced after the collapse of a massive star \citep{Woosley:1993ga}. The nature of short-duration GRBs was under debate for many decades until the first detection of a neutron star merger through gravitational waves and observations of its electromagnetic counterpart \citep{Blinnikov:1984no,Goodman:1986ar,Abbott:2017gr}.

GRBs show two different phases in their light curves since the outburst: the prompt phase and the afterglow. The prompt phase of a GRB happens during the first few seconds after the explosion, showing great variability in the light curves. This short period of time is the most luminous of the explosion. The afterglow, on the other hand, is the time period after the prompt emission in which the ejected material interacts with the surrounding environment. This phase may last for years. Great progress has been made in the last decades trying to understand both phases \citep[e.g.,][]{Meszaros:1994re,Sari:1998pi}.

\begin{figure}
  \centering
  \includegraphics[width=0.9\columnwidth]{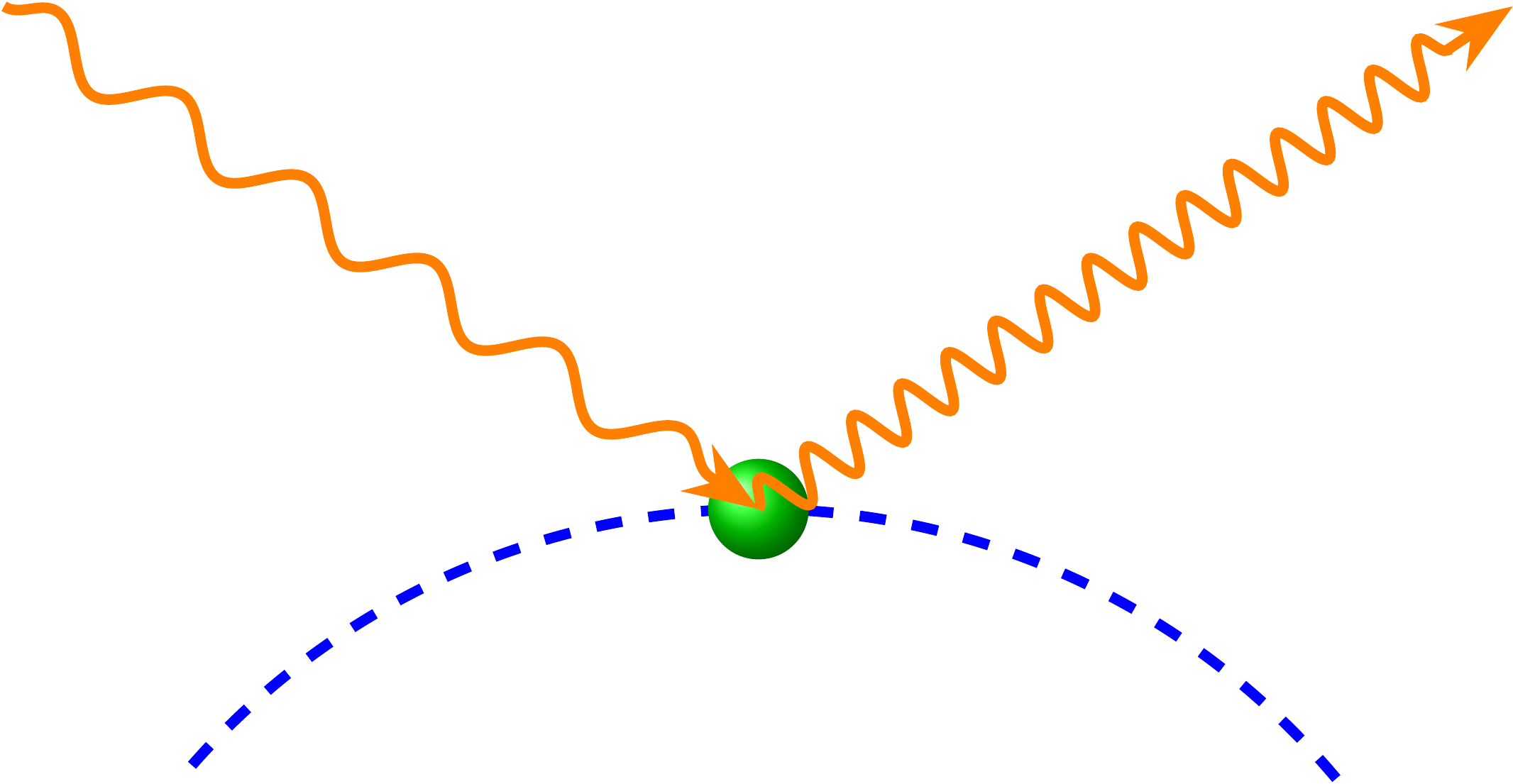}
  \caption{Inverse Compton scattering}
\label{fig:IC}
\end{figure}

The inverse Compton (IC) scattering is a radiative process in which an ultra-relativistic charged particle (e.g., an electron, see Fig.~\ref{fig:IC}), interacts with a low frequency photon. The electron transfers part of its energy to the photon, giving way to a photon with higher frequency. Theoretical models have been developed considering this process since it may account for the highly energetic signal observed from GRBs, reaching MeV or even TeV. In recent decades, studies have been made of the role that this radiative process plays in the evolution of accelerated particles \citep{BarniolDuran:2012bo,Panaitescu:2019ad}. In this radiative process, there is a regime of energies beyond which the classical Thomson cross section no longer applies and quantum corrections need to be made. This is the so-called Klein-Nishina (KN) regime. Being in the KN regime also implies corrections to the radiative cooling, whose full mathematical expression is highly non-linear. Analytical approximations have been developed in order to account for these particle energy losses in relativistic outflows \citep{Moderski:2005si,Nakar:2009an}, and have been applied to GRBs \citep{BarniolDuran:2012bo,Panaitescu:2019ad}.

In order to study the effects of KN cooling, a numerical approach to these corrections has been developed in  order to simulate the evolution of a distribution of electrons interacting with different kinds of external radiation fields. The present work is structured as follows: in Sec.~\ref{sec:numerics} we will describe the numerical methods and model employed to describe GRB afterglows. In Sec.~\ref{sec:num-KN} we show the numerical implementation of the KN cooling, and the corresponding qualitative tests. In Sec.~\ref{sec:application} we present the application of the method to GRB190114C, and in Sec.~\ref{sec:conclu} we present our conclusions.

\section{The Numerical Method}\label{sec:numerics}

In the present section we will describe the model and numerical methods employed to describe GRB afterglows. We also describe the implementation of the KN cooling.

\subsection{The kinetic equation}\label{sec:kinetic}

The numerical code \texttt{Paramo} \citep{RuedaBecerril:2020Paramo} has been developed to monitor the evolution of a distribution of ultra-relativistic particles, and their radiative signature for different scenarios. The evolution of a distribution of relativistic particles $n'(\gamma', t')$, with $\gamma'$ being the particle Lorentz factor and $t'$ the time variable\footnote{Primed quantities denote those measured in the comoving frame of the fluid, while unprimed ones denote those measured in the rest frame of the central engine, also denoted as the ``lab frame''.}, is given by the kinetic equation:
\begin{align}
  \dfrac{\partial n'(\gamma', t')}{\partial t'} = & \frac{1}{2} \dfrac{\partial^2}{\partial\gamma'^2} \left[ D'(\gamma', t') n'(\gamma', t') \right] \nonumber \\
    & - \dfrac{\partial}{\partial\gamma'} \left[ \dot{\gamma}'(\gamma', t') n'(\gamma', t') \right] + Q'(\gamma', t') - \dfrac{n'(\gamma', t')}{t'_{\mathrm{esc}}}, \label{eq:FP}
\end{align}
better known as the Fokker-Planck equation. The Fokker-Planck equation is a diffusion equation in the momentum space which describes the dynamics of $n'$ under diffusion processes ($D'$) and energy gains/losses ($\dot{\gamma}'$). The general form includes a source term which accounts for the injection of particles to the system ($Q'$), and also a sink term that accounts for particle losses, e.g., that after some period of time $t'_{\mathrm{esc}}$ particles leave the system. The solution of this equation will give us the profile of $n'$ for a given $t'$. In the code \texttt{Paramo} \citep{RuedaBecerril:2020Paramo}, robust numerical methods have been implemented to solve the Fokker-Plank equation \citep{Park:1996pe}.

\subsection{The fireball model}\label{sec:blaswave}

GRB afterglows have been modelled as radiation produced by charged particles accelerated at the the shock front of relativistic blast waves \citep{Paczynski:1993rh}. In this model, an amount of energy $E_0$ is injected into a sphere with radius $r_0$ with initial rest mas $M_0$. This ``fireball'' evolves into a thin shell with initial bulk Lorentz factor $\Gamma_0$, propagating through the circumburst medium with uniform rest mass density $\rho_0$. The blast wave will sweep up matter as it expands, decelerating the shell and reheating the material in it. Following \citet{Petropoulou:2009ma}, the bulk Lorentz factor for a non-radiative blast wave is given by
\begin{equation}\label{eq:Gamma}
  \Gamma(r) = \dfrac{\lambda \left(x^3 - 1\right) + \Gamma_0}{{\left(1 + 2 \Gamma_0 \lambda \left(x^3 - 1\right) + \lambda^2 {\left(x^3 - 1\right)}^2 \right)}^{1/2}},
\end{equation}
where $r$ is the radius of the blast wave, $\lambda \equiv 4 \pi \rho_0 r_0^3 / 3 M_0$, $x \equiv r / r_0$, and $M_0 \equiv E_0 / \Gamma_0 c^2$.

\subsubsection{Injection of particles}

Through the heating process, particles moving with the shell may be accelerated to relativistic energies. We assume that a fraction $\epsilon_{e}$ of the accreted kinetic energy goes to electrons, accelerating them into a power-law distribution. We monitor these particles by solving Eq.~\eqref{eq:FP} numerically, as described above. We assume a continuous injection of a power-law distribution of particles; i.e., 
\begin{equation}
  Q'(\gamma', t') = Q_0 {\gamma'}^{-p} H\left(\gamma'; \gamma_{\min}, \gamma_{\max}\right), \mbox { for } t > 0,
\end{equation}
where $Q_0$ is the normalization coefficient, $p$ is the power-law index and $H(x; a, b) = 1$ if $a \leq x \leq b$, and 0 otherwise.

\subsubsection{Magnetic field}\label{sec:magnetic}

Assuming that a fraction of the shock energy $\epsilon_B$ transforms into magnetic energy \citep[e.g.,][]{Sari:1998pi}, we get that the magnetic field strength, in the comoving frame, is given by
\begin{equation}
  B'(r) = {\left( 32 \pi n_0 m_p \epsilon_B \right)}^{1 / 2} c \Gamma(r),
\end{equation}
where $m_p$ is the proton mass and $n_0 \equiv \rho_0 / m_p$.

\subsubsection{Evolution}

Our electrons and their emitted radiation are monitored in the comoving frame. The proper transformations are performed in order to get the radiation flux in the reference frame of an observer on Earth. However, the evolution of the blast wave is being held in the lab frame. The time in the comoving frame of the evolution of the shock is given by
\begin{equation}\label{eq:t-com}
  t' = \int_{0}^{t} \dfrac{dt}{\Gamma} = \int_{r_0}^{r} \dfrac{dr}{\Gamma \beta c},
\end{equation}
where $\beta = {(1 - \Gamma^{-2})}^{1 / 2}$. The integral in Eq.~\eqref{eq:t-com} is performed numerically to obtain the time and time-step at each point of the evolution of the blast wave. On the other hand, the time in the observer frame is given by
\begin{equation}\label{eq:t-obs}
  t_{\mathrm{obs}} = (1 + z) \int_{r_0}^{r} \dfrac{dr}{\mathcal{D} \Gamma \beta c}
\end{equation}
where $z$ is the redshift and $\mathcal{D} = 1 / (1 - \beta \cos\theta_{\mathrm{obs}}) \Gamma$ is the Doppler factor, with $\theta_{\mathrm{obs}}$ being the observer viewing angle with respect to the propagation axis of the outflow. The integral in Eq.~\eqref{eq:t-obs} is also performed numerically.

\subsection{Numerical synchrotron and IC}\label{sec:syn-ic}

We assume that the emission that an earthly observer receives, comes from a region of the blast wave which in the comoving frame can be approximated as a blob of radius $R'_\mathrm{b} = r / \Gamma$. The electrons in it emit isotropically. For this geometry the energy flux is given by \citep{Gould:1979cs}
\begin{equation}\label{eq:nuFnu}
  \nu_\mathrm{obs} F_{\nu_\mathrm{obs}} = \dfrac{1 + z}{d_\mathrm{L}^2} f(\tau_{\nu'}') \mathcal{D}^{4} V' \nu' j_{\nu'}',
\end{equation}
where $d_\mathrm{L}$ is the luminosity distance, $\nu_\mathrm{obs} = \mathcal{D} \nu' / (1 + z)$,  $\tau_{\nu'}' \equiv 2 R_{\rm b}' \kappa'_{\nu'}$, with $j_{\nu'}'$ and $\kappa_{\nu'}'$ being the synchrotron emissivity and self-absorption \citep{Rybicki:1979}, respectively, and  \citep{Gould:1979cs,Dermer:2009}
\begin{equation}
  f(\tau) \equiv \dfrac{3}{\tau} \left( \dfrac{1}{2} + \dfrac{\exp(-\tau)}{\tau} - \dfrac{1 - \exp(-\tau)}{\tau^2} \right),
\end{equation}
is the optical depth function for a spherical blob.

The total emissivity $j'_{\nu'}$ and absorption $\kappa'_{\nu'}$ of a distribution of particles is computed in \texttt{Paramo} employing the numerical techniques developed in \citet{RuedaBecerril:2017phd,RuedaBecerril:2020kn}. The main radiative processes we focus on are: synchrotron and IC (SSC and EIC). For the synchrotron process, self-absorption is taken into account.

\section{Numerical Klein-Nishina}\label{sec:num-KN}

The IC process basically consists on the energy transfer from an ultra-relativistic electron to a low-frequency photon. This implies a loss of energy, or ``cooling'', of the electron. This radiative cooling, in the classical Thomson regime, is given by \citep{Rybicki:1979}
\begin{equation}\label{eq:dotg-Thom}
  \dot{\gamma}'_\mathrm{rad} = \dfrac{4 \sigma_\mathrm{T} c \beta_e'^2 \gamma'^2 u_0'}{3 m_e c^2},
\end{equation}
where $\sigma_\mathrm{T}$ is the Thomson cross-section, $\gamma'$ the electron Lorentz factor, $\beta'_e = {(1 - \gamma'^{-2})}^{1 / 2}$, and
\begin{equation}\label{eq:urad-Thom}
  u_0' = \int d\nu'_0 u_{\nu'_0}',
\end{equation}
is the total energy density of the incoming radiation field, with $\nu_0'$ and $u_{\nu_0'}'$ the frequency and spectral energy density of the incoming radiation field.

Relations \eqref{eq:dotg-Thom} and \eqref{eq:urad-Thom} are only valid if $\gamma' h \nu' \ll m_e c^2$. When $\gamma' h \nu' > m_e c^2$, the IC scattering is now described by the KN cross-section; i.e.,
\begin{equation}\label{eq:dotg-KN}
  \dot{\gamma}_\mathrm{KN} = \dfrac{4 c \sigma_\mathrm{KN} \gamma'^2 u'_0}{3 m_e c^2},
\end{equation}
where
\begin{equation}
  \sigma_\mathrm{KN} = \dfrac{\sigma_\mathrm{T}}{u'_0} \int d\varepsilon'_0 u'_{\varepsilon'_0} \dfrac{9 g(b)}{b^3}
\end{equation}
is the KN cross-section, with $\varepsilon'_0 \equiv h \nu'_0 / m_e c^2$, $b \equiv 4 \gamma' \varepsilon'_0 $, and  \citep[e.g.,][]{Moderski:2005si}
\begin{align}
  g(b) = & \left( \dfrac{b}{2} + 6 + \dfrac{6}{b} \right) \ln(1 + b) - \left( \dfrac{11 b^3}{12} + 6 b^2 + 9 b + 4 \right) \nonumber \\
  & \times \dfrac{1}{{(1 + b)}^2} - 2 + 2 \mathrm{Li}_2(-b), \label{eq:gKN}
\end{align}
where Li$_2(x)$ is the dilogarithm function. The above shows that the KN cooling is highly nonlinear, and difficult to calculate in dynamical systems.

A numerical tool has been developed, and implemented in the code \texttt{Paramo}, capable of calculating the radiative energy losses of charged particles, taking into account the KN corrections mentioned above, for radiation fields with monochromatic and arbitrary energy density profiles. The numerical tool consists in taking the asymptotic approximations of Eq.~\eqref{eq:gKN} and patching them with polynomial approximations \citep[see][for an application of this technique to synchrotron emission]{RuedaBecerril:2017phd}.

\begin{figure}
  \includegraphics[width=\linewidth]{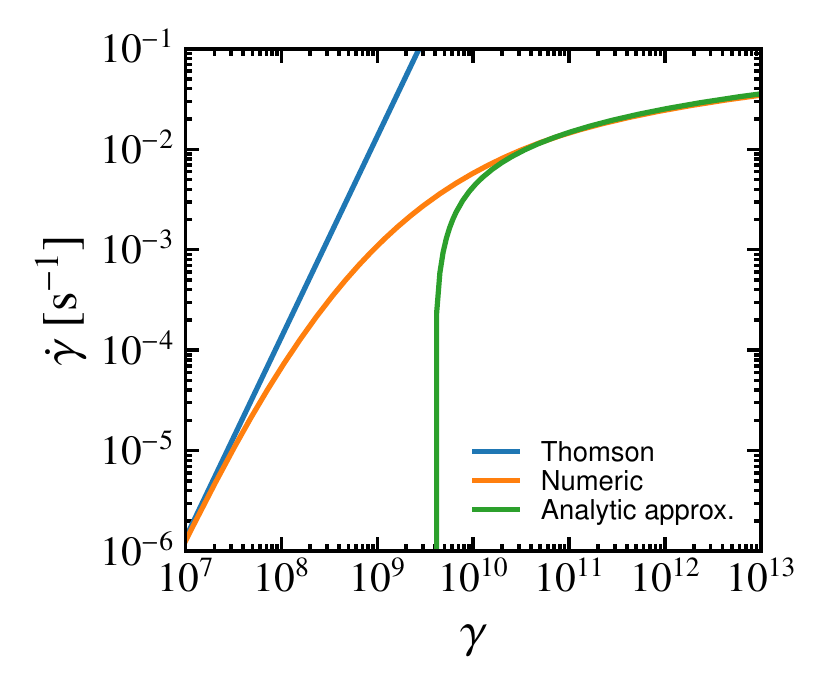}
  \caption{Radiative cooling due to a blackbody radiation field. Blue, orange and green lines correspond to the Thomson, numerical and asymptotic approximations, respectively.}
  \label{fig:KN-bb}
\end{figure}

In Fig.~\ref{fig:KN-bb} we show a qualitative comparison of black body radiative cooling. The Thomson regime appears as a blue line, while the green line corresponds to Eq.~(6.54) in \citet{Dermer:2009}. The numerical radiative cooling is shown as an orange line. As we can see, the analytic approximation shows a drop-off at $\gamma' \sim 4 \times 10^9$, producing a large gap between the KN and Thomson regimes. Our numerical approach matches both the deep KN and Thomson regimes with a smooth transition between them.

\begin{figure}
  \includegraphics[width=\linewidth]{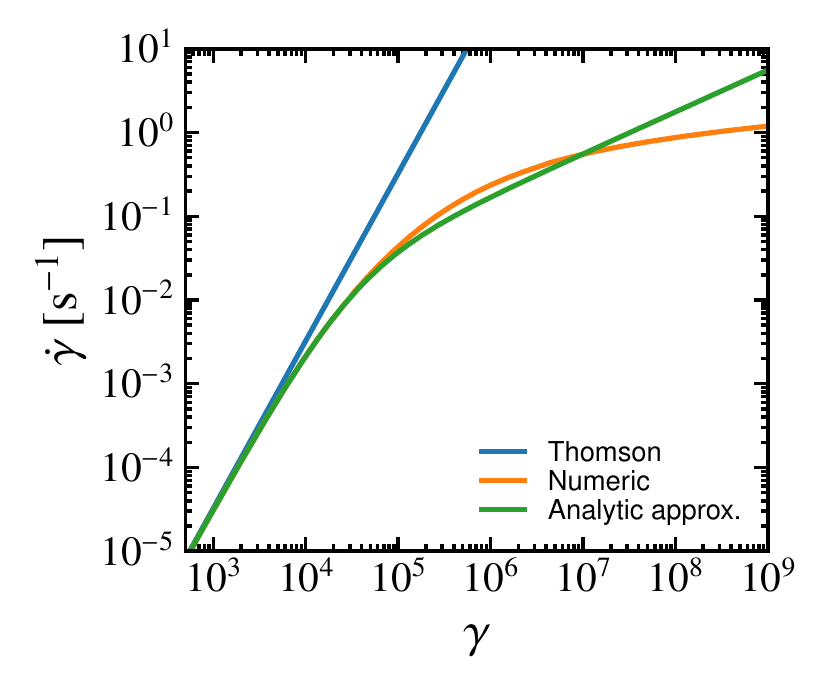}
  \caption{Radiative cooling due to a monochromatic radiation field. Blue, orange and green lines correspond to the Thomson, numerical and asymptotic approximations, respectively.}
  \label{fig:KN-mono}
\end{figure}

In Fig.~\ref{fig:KN-mono} we show a qualitative comparison of radiative cooling coefficients for the case of a monochromatic radiation field. In the same manner as in Fig.~\ref{fig:KN-bb}, the blue line corresponds to the Thomson regime, the orange line the numerical approximation, and the green line corresponds to the analytic approximation presented in \citet{Moderski:2005si}. As mentioned in that paper, such an approximation is valid for $\gamma' \lesssim 10^{4}$, which corresponds to the transition from the Thomson to the KN regime. Our numerical approach matches both the Thomson and trans-KN regimes. For large Lorentz factors, however, our numerical calculations diverge from the analytic approximation. The physical implication of this divergence is that with the analytical approximation, ultra-relativistic particles will cool down more efficiently than with the numerical approximation.

\section{Application to GRB190114C}\label{sec:application}

The workhorse model for GRB explosions over the years has been the fireball model \citep{Cavallo:1978re,Paczynski:1986ga,Rees:1992me}. As we have described in Sec.~\ref{sec:blaswave}, this model assumes that a blast-wave is formed at the moment of the explosion. The relativistic outflow propagates through the circumburst medium, accreting matter on its way. A fraction of the energy of the accreted matter is transferred to charged particles in the plasma (e.g., electrons), accelerating them to ultrarelativistic speeds. These particles interact with the magnetic field enhanced by the shock (see Sec.~\ref{sec:magnetic}), emitting synchrotron radiation. These synchrotron photons produced \textit{in situ} may interact via IC scattering with the same electrons that have emitted them, producing what is called the synchrotron self-Compton (SSC) radiation. In the same manner, when the blast wave reaches the circumburst environment the accelerated electrons at the shock front may interact also via IC with the photons in the surrounding environment, contributing to the high-energy spectral signal with an external inverse Compton (EIC) component. The radiation field assumed for the EIC is monochromatic with frequency $\nu_0$ with energy density $u_0$. In the comoving frame of the outflow, $\nu'_0 = \Gamma \nu_0$ and $u'_0 = \Gamma^2 (1 + \beta^2 / 3) u_0$.

\begin{figure}
  \includegraphics[width=\columnwidth]{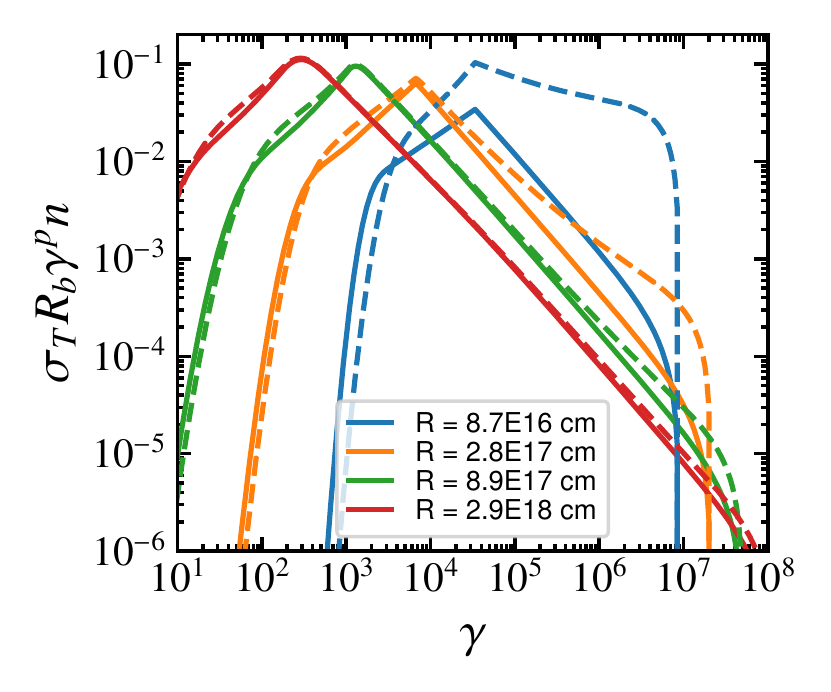}
  \caption{Snapshots of the EED at the shock front of GRB190114C, at different radii from the location of the explosion.}
  \label{fig:EED}
\end{figure}

\begin{table}
  \centering
  \begin{tabular}{cc}
    \toprule
    Parameter & Valor \\
    \midrule
    $d_{\mathrm{L}}$ & $2.3 \times 10^9$ pc\\
    $z$ & 0.4245\\
    $p$ & 2.6\\
    $\epsilon_B$ & $8 \times 10^{-5}$\\
    $\epsilon_e$ & 0.07\\
    $E_0$ & $8 \times 10^{53}$ erg\\
    $\Gamma_0$ & 700\\
    $n_{0}$ & 0.5 cm$^{-3}$\\
    $u_{0}$ & $7.5 \times 10^{-9}$ erg cm$^{-3}$\\
    $\nu_{0}$ & 0.02 eV$/h$\\
    $\theta_{\mathrm{obs}}$ & $0^{\circ}$\\
    \bottomrule
  \end{tabular}
  \caption{Initial conditions for the fireball model. The values were taken from the best fit for GRB190114C as reported in \citet{Acciari:2019bh}. The parameters for the external photon field ($u_0$ and $\nu_0$) correspond to those values employed by \citet{Zhang:2020ch} for the same GRB.}
\label{tab:params}
\end{table}

In order to test the numerical KN in an evolutionary scenario, we have implemented the self-similar solution for the evolution of a relativistic blast-wave \citep{Blandford:1976mc}, in the way described in Sec.~\ref{sec:blaswave}, which recreates the relativistic outflow in the fireball model for GRBs. This solution will give us the position $R$ and bulk Lorentz factor $\Gamma$ of the shock wave. Assuming that the radiating particles are electrons at the shock front of the explosion, accelerated into a power-law distribution with index $p$, and ignoring any diffusion (i.e., $D' = 0$), one can solve the Fokker-Planck equation and get the profile of the electron energy distribution (EED) at each radius of the explosion. We have taken the fit to GRB190114C reported by \citet{Acciari:2019bh} as initial conditions for our simulation (see Tab.~\ref{tab:params}). The photon fields considered to be interacting with the electrons are: synchrotron and external photons (see Tab.~\ref{tab:params}).
\begin{figure*}
  \includegraphics[width=0.49\linewidth]{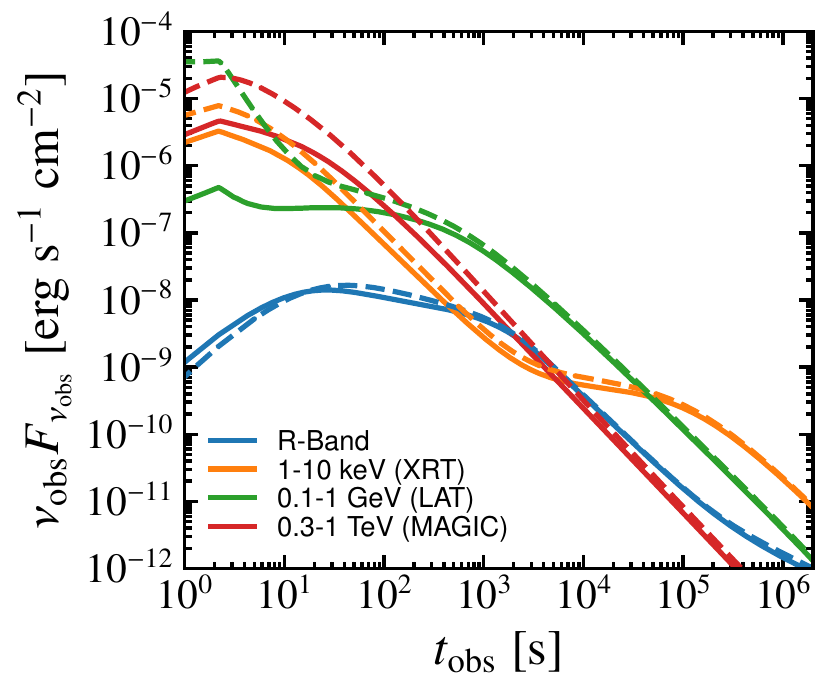}
  \hfill
  \includegraphics[width=0.49\linewidth]{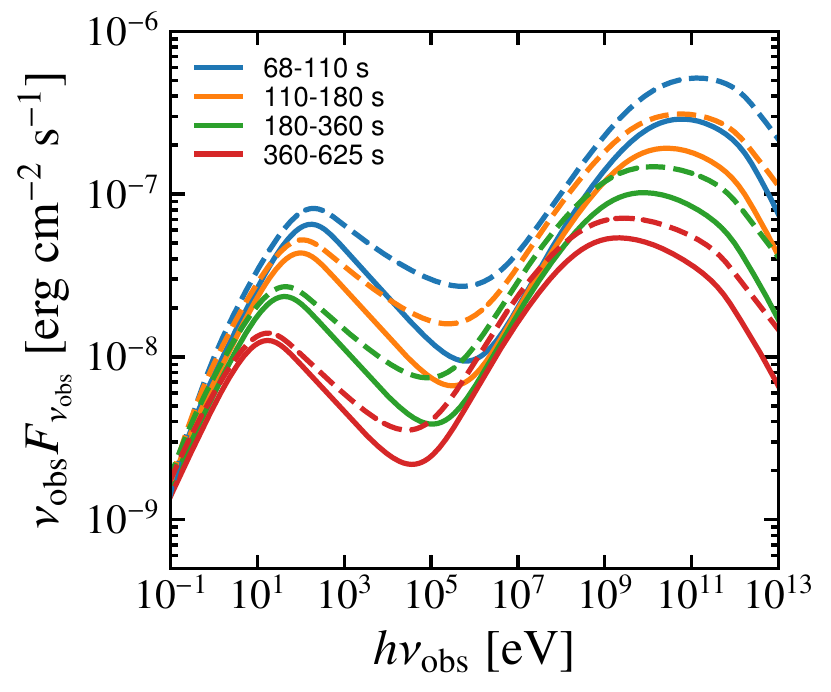}
  \caption{\textit{Left Panel}: Light curves (cf. Fig~1 in \citet{Acciari:2019bh}). \textit{Right Panel}: Averaged SEDs (cf. Fig~2 in \citet{Acciari:2019bh}).}
\label{fig:spec}
\end{figure*}

In Fig.~\ref{fig:EED} we show the EEDs at different radii of the blast-wave. In solid lines we show the simulations with pure Thomson cooling, and in dashed lines those with the KN cooling turned on. During the first stages of the evolution of the blast-wave, high-energy electrons accumulate when the KN cooling is taken into account. This is due to the fact that radiative cooling is less efficient when it enters into the KN regime, occurring mostly at the high-energy end, allowing particles to pile-up and harden the energy index. On the other hand, at later stages, the KN cooling is dominated by synchrotron and radiative cooling in the Thomson regime.

In Fig.~\ref{fig:spec} we show the light curves for different wavebands (left panel) and averaged spectral energy distributions (SEDs) for different time intervals (right panel). In the same manner as in Fig.~\ref{fig:EED}, solid lines we show the simulations with radiative cooling in the pure Thomson regime, while the dashed lines belong to the simulations with the numerical KN cooling. We can see that the KN cooling effects are evident at the early stages of the explosion. At later stages (i.e., larger radius), we can see that the EEDs and light curves are similar for both coolings. This would mean that the accelerated electrons are interacting with the surrounding photons in the Thomson regime. Through the SEDs we can appreciate how the IC component (SSC+EIC) is $\sim$twice louder during the first 100s, and has a harder spectrum.

\section{Conclusions}\label{sec:conclu}

In this work we present a new numerical tool able to calculate the radiative KN cooling and implement it to relativistic outflows like GRBs. In this work we have found in particular that:
\begin{itemize}
  \item The numerical KN tool has been compared and contrasted with analytical approximations, showing a smooth and consistent transition from the Thomson regime, through the trans-KN, and all the way to the deep KN regimes.
  \item The numerical code \texttt{Paramo}, together with the numerical KN, has been tested in an evolutionary scenario such as the fireball model of GRBs.
  \item The KN cooling has shown to be inefficient in cooling ultrarelativistic electrons, allowing them to pile-up in the first stages of the blast-wave. These un-cooled high-energy electrons will contribute in emitting more synchrotron, SSC and EIC emission during the first $\sim 100$~s after the explosion.
  \item During most of the afterglow stage of a GRB, radiative cooling is purely Thomson.
\end{itemize}

\section*{Acknowledgments}

JMRB acknowledges the support from the Mexican National Council of Science and Technology (\fundingAgency{CONACYT}) with the Post-doctoral Fellowship under the program Post-doctoral Stays Abroad \fundingNumber{CVU332030}.

\bibliography{Becerril}%

\begin{thebibliography}{}

\bibitem [\protect \citeauthoryear {%
Abbott~et al.%
}{%
Abbott~et al.%
}{%
{\protect \APACyear {2017}}%
}]{%
Abbott:2017gr}
\APACinsertmetastar {%
Abbott:2017gr}%
\begin{APACrefauthors}%
Abbott~et al., B\BPBI P.%
\end{APACrefauthors}%
\unskip\
\newblock
\APACrefYearMonthDay{2017}{oct}{},
\newblock
\unskip
\newblock
\APACjournalVolNumPages{Astrophys. J.}{\bf 848}{2}{L13}.
\newblock
\begin{APACrefDOI} \doi{10.3847/2041-8213/aa920c} \end{APACrefDOI}
\PrintBackRefs{\CurrentBib}

\bibitem [\protect \citeauthoryear {%
Acciari%
\ \protect \BOthers {.}}{%
Acciari%
\ \protect \BOthers {.}}{%
{\protect \APACyear {2019}}%
}]{%
Acciari:2019bh}
\APACinsertmetastar {%
Acciari:2019bh}%
\begin{APACrefauthors}%
Acciari, V\BPBI A.%
, Ansoldi, S.%
, Antonelli, L\BPBI A.%
\ et al.\end{APACrefauthors}%
\unskip\
\newblock
\APACrefYearMonthDay{2019}{nov}{},
\newblock
\unskip
\newblock
\APACjournalVolNumPages{Nature}{\bf 575}{7783}{459--463}.
\newblock
\begin{APACrefDOI} \doi{10.1038/s41586-019-1754-6} \end{APACrefDOI}
\PrintBackRefs{\CurrentBib}

\bibitem [\protect \citeauthoryear {%
{Barniol Duran}%
, Bo{\v{s}}njak%
\BCBL {}\ \BBA {} Kumar%
}{%
{Barniol Duran}%
\ \protect \BOthers {.}}{%
{\protect \APACyear {2012}}%
}]{%
BarniolDuran:2012bo}
\APACinsertmetastar {%
BarniolDuran:2012bo}%
\begin{APACrefauthors}%
{Barniol Duran}, R.%
, Bo{\v{s}}njak, {\v{Z}}.%
\BCBL {}\ \BBA {} Kumar, P.%
\end{APACrefauthors}%
\unskip\
\newblock
\APACrefYearMonthDay{2012}{aug}{},
\newblock
\unskip
\newblock
\APACjournalVolNumPages{Mon. Not. R. Astron. Soc.}{\bf 424}{4}{3192--3200}.
\newblock
\begin{APACrefDOI} \doi{10.1111/j.1365-2966.2012.21533.x} \end{APACrefDOI}
\PrintBackRefs{\CurrentBib}

\bibitem [\protect \citeauthoryear {%
Blandford%
\ \BBA {} McKee%
}{%
Blandford%
\ \BBA {} McKee%
}{%
{\protect \APACyear {1976}}%
}]{%
Blandford:1976mc}
\APACinsertmetastar {%
Blandford:1976mc}%
\begin{APACrefauthors}%
Blandford, R\BPBI D.%
\BCBT {}\ \BBA {} McKee, C\BPBI F.%
\end{APACrefauthors}%
\unskip\
\newblock
\APACrefYearMonthDay{1976}{}{},
\newblock
\unskip
\newblock
\APACjournalVolNumPages{Phys. Fluids}{\bf 19}{8}{1130--1138}.
\newblock
\begin{APACrefDOI} \doi{10.1063/1.861619} \end{APACrefDOI}
\PrintBackRefs{\CurrentBib}

\bibitem [\protect \citeauthoryear {%
{Blinnikov}%
\ \BBA {} et al.%
}{%
{Blinnikov}%
\ \BBA {} et al.%
}{%
{\protect \APACyear {1984}}%
}]{%
Blinnikov:1984no}
\APACinsertmetastar {%
Blinnikov:1984no}%
\begin{APACrefauthors}%
{Blinnikov}, S\BPBI I.%
\BCBT {}\ \BBA {} et al.%
\end{APACrefauthors}%
\unskip\
\newblock
\APACrefYearMonthDay{1984}{{\APACmonth{04}}}{},
\newblock
\unskip
\newblock
\APACjournalVolNumPages{Sov. Astron. Let.}{\bf 10}{}{177-179}.
\PrintBackRefs{\CurrentBib}

\bibitem [\protect \citeauthoryear {%
Cavallo%
\ \BBA {} Rees%
}{%
Cavallo%
\ \BBA {} Rees%
}{%
{\protect \APACyear {1978}}%
}]{%
Cavallo:1978re}
\APACinsertmetastar {%
Cavallo:1978re}%
\begin{APACrefauthors}%
Cavallo, G.%
\BCBT {}\ \BBA {} Rees, M\BPBI J.%
\end{APACrefauthors}%
\unskip\
\newblock
\APACrefYearMonthDay{1978}{jul}{},
\newblock
\unskip
\newblock
\APACjournalVolNumPages{Mon. Not. R. Astron. Soc.}{\bf 183}{3}{359--365}.
\newblock
\begin{APACrefDOI} \doi{10.1093/mnras/183.3.359} \end{APACrefDOI}
\PrintBackRefs{\CurrentBib}

\bibitem [\protect \citeauthoryear {%
{Dermer}%
\ \BBA {} {Menon}%
}{%
{Dermer}%
\ \BBA {} {Menon}%
}{%
{\protect \APACyear {2009}}%
}]{%
Dermer:2009}
\APACinsertmetastar {%
Dermer:2009}%
\begin{APACrefauthors}%
{Dermer}, C\BPBI D.%
\BCBT {}\ \BBA {} {Menon}, G.%
\end{APACrefauthors}%
\unskip\
\newblock
\APACrefYear{2009},
\newblock
\APACrefbtitle {{High Energy Radiation from Black Holes. Gamma Rays, Cosmic
  Rays, and Neutrinos}} {{High Energy Radiation from Black Holes. Gamma Rays,
  Cosmic Rays, and Neutrinos}}.
\newblock
\APACaddressPublisher{Princeton}{Princeton University Press}.
\PrintBackRefs{\CurrentBib}

\bibitem [\protect \citeauthoryear {%
Goodman%
}{%
Goodman%
}{%
{\protect \APACyear {1986}}%
}]{%
Goodman:1986ar}
\APACinsertmetastar {%
Goodman:1986ar}%
\begin{APACrefauthors}%
Goodman, J.%
\end{APACrefauthors}%
\unskip\
\newblock
\APACrefYearMonthDay{1986}{sep}{},
\newblock
\unskip
\newblock
\APACjournalVolNumPages{Astrophys. J.}{\bf 308}{}{L47}.
\newblock
\begin{APACrefDOI} \doi{10.1086/184741} \end{APACrefDOI}
\PrintBackRefs{\CurrentBib}

\bibitem [\protect \citeauthoryear {%
{Gould}%
}{%
{Gould}%
}{%
{\protect \APACyear {1979}}%
}]{%
Gould:1979cs}
\APACinsertmetastar {%
Gould:1979cs}%
\begin{APACrefauthors}%
{Gould}, R\BPBI J.%
\end{APACrefauthors}%
\unskip\
\newblock
\APACrefYearMonthDay{1979}{7}{},
\newblock
\unskip
\newblock
\APACjournalVolNumPages{\aap}{\bf 76}{3}{306--311}.
\PrintBackRefs{\CurrentBib}

\bibitem [\protect \citeauthoryear {%
Meszaros%
, Rees%
\BCBL {}\ \BBA {} Papathanassiou%
}{%
Meszaros%
\ \protect \BOthers {.}}{%
{\protect \APACyear {1994}}%
}]{%
Meszaros:1994re}
\APACinsertmetastar {%
Meszaros:1994re}%
\begin{APACrefauthors}%
Meszaros, P.%
, Rees, M\BPBI J.%
\BCBL {}\ \BBA {} Papathanassiou, H.%
\end{APACrefauthors}%
\unskip\
\newblock
\APACrefYearMonthDay{1994}{nov}{},
\newblock
\unskip
\newblock
\APACjournalVolNumPages{Astrophys. J.}{\bf 432}{}{181}.
\newblock
\begin{APACrefDOI} \doi{10.1086/174559} \end{APACrefDOI}
\PrintBackRefs{\CurrentBib}

\bibitem [\protect \citeauthoryear {%
Moderski%
, Sikora%
, Coppi%
\BCBL {}\ \BBA {} Aharonian%
}{%
Moderski%
\ \protect \BOthers {.}}{%
{\protect \APACyear {2005}}%
}]{%
Moderski:2005si}
\APACinsertmetastar {%
Moderski:2005si}%
\begin{APACrefauthors}%
Moderski, R.%
, Sikora, M.%
, Coppi, P\BPBI S.%
\BCBL {}\ \BBA {} Aharonian, F.%
\end{APACrefauthors}%
\unskip\
\newblock
\APACrefYearMonthDay{2005}{nov}{},
\newblock
\unskip
\newblock
\APACjournalVolNumPages{Mon. Not. R. Astron. Soc.}{\bf 363}{3}{954--966}.
\newblock
\begin{APACrefDOI} \doi{10.1111/j.1365-2966.2005.09494.x} \end{APACrefDOI}
\PrintBackRefs{\CurrentBib}

\bibitem [\protect \citeauthoryear {%
Nakar%
, Ando%
\BCBL {}\ \BBA {} Sari%
}{%
Nakar%
\ \protect \BOthers {.}}{%
{\protect \APACyear {2009}}%
}]{%
Nakar:2009an}
\APACinsertmetastar {%
Nakar:2009an}%
\begin{APACrefauthors}%
Nakar, E.%
, Ando, S.%
\BCBL {}\ \BBA {} Sari, R.%
\end{APACrefauthors}%
\unskip\
\newblock
\APACrefYearMonthDay{2009}{sep}{},
\newblock
\unskip
\newblock
\APACjournalVolNumPages{Astrophys. J.}{\bf 703}{1}{675--691}.
\newblock
\begin{APACrefDOI} \doi{10.1088/0004-637X/703/1/675} \end{APACrefDOI}
\PrintBackRefs{\CurrentBib}

\bibitem [\protect \citeauthoryear {%
Paczynski%
}{%
Paczynski%
}{%
{\protect \APACyear {1986}}%
}]{%
Paczynski:1986ga}
\APACinsertmetastar {%
Paczynski:1986ga}%
\begin{APACrefauthors}%
Paczynski, B.%
\end{APACrefauthors}%
\unskip\
\newblock
\APACrefYearMonthDay{1986}{sep}{},
\newblock
\unskip
\newblock
\APACjournalVolNumPages{Astrophys. J.}{\bf 308}{}{L43}.
\newblock
\begin{APACrefDOI} \doi{10.1086/184740} \end{APACrefDOI}
\PrintBackRefs{\CurrentBib}

\bibitem [\protect \citeauthoryear {%
{Paczynski}%
\ \BBA {} {Rhoads}%
}{%
{Paczynski}%
\ \BBA {} {Rhoads}%
}{%
{\protect \APACyear {1993}}%
}]{%
Paczynski:1993rh}
\APACinsertmetastar {%
Paczynski:1993rh}%
\begin{APACrefauthors}%
{Paczynski}, B.%
\BCBT {}\ \BBA {} {Rhoads}, J\BPBI E.%
\end{APACrefauthors}%
\unskip\
\newblock
\APACrefYearMonthDay{1993}{{\APACmonth{11}}}{},
\newblock
\unskip
\newblock
\APACjournalVolNumPages{\apjl}{\bf 418}{}{L5}.
\newblock
\begin{APACrefDOI} \doi{10.1086/187102} \end{APACrefDOI}
\PrintBackRefs{\CurrentBib}

\bibitem [\protect \citeauthoryear {%
Panaitescu%
}{%
Panaitescu%
}{%
{\protect \APACyear {2019}}%
}]{%
Panaitescu:2019ad}
\APACinsertmetastar {%
Panaitescu:2019ad}%
\begin{APACrefauthors}%
Panaitescu, A.%
\end{APACrefauthors}%
\unskip\
\newblock
\APACrefYearMonthDay{2019}{nov}{},
\newblock
\unskip
\newblock
\APACjournalVolNumPages{Astrophys. J.}{\bf 886}{2}{106}.
\newblock
\begin{APACrefDOI} \doi{10.3847/1538-4357/ab4e17} \end{APACrefDOI}
\PrintBackRefs{\CurrentBib}

\bibitem [\protect \citeauthoryear {%
Park%
\ \BBA {} Petrosian%
}{%
Park%
\ \BBA {} Petrosian%
}{%
{\protect \APACyear {1996}}%
}]{%
Park:1996pe}
\APACinsertmetastar {%
Park:1996pe}%
\begin{APACrefauthors}%
Park, B\BPBI T.%
\BCBT {}\ \BBA {} Petrosian, V.%
\end{APACrefauthors}%
\unskip\
\newblock
\APACrefYearMonthDay{1996}{mar}{},
\newblock
\unskip
\newblock
\APACjournalVolNumPages{Astrophys. J. Suppl. Ser.}{\bf 103}{}{255}.
\newblock
\begin{APACrefDOI} \doi{10.1086/192278} \end{APACrefDOI}
\PrintBackRefs{\CurrentBib}

\bibitem [\protect \citeauthoryear {%
Petropoulou%
\ \BBA {} Mastichiadis%
}{%
Petropoulou%
\ \BBA {} Mastichiadis%
}{%
{\protect \APACyear {2009}}%
}]{%
Petropoulou:2009ma}
\APACinsertmetastar {%
Petropoulou:2009ma}%
\begin{APACrefauthors}%
Petropoulou, M.%
\BCBT {}\ \BBA {} Mastichiadis, A.%
\end{APACrefauthors}%
\unskip\
\newblock
\APACrefYearMonthDay{2009}{nov}{},
\newblock
\unskip
\newblock
\APACjournalVolNumPages{Astron. Astrophys.}{\bf 507}{2}{599--610}.
\newblock
\begin{APACrefDOI} \doi{10.1051/0004-6361/200912970} \end{APACrefDOI}
\PrintBackRefs{\CurrentBib}

\bibitem [\protect \citeauthoryear {%
Rees%
\ \BBA {} M{\'{e}}sz{\'{a}}ros%
}{%
Rees%
\ \BBA {} M{\'{e}}sz{\'{a}}ros%
}{%
{\protect \APACyear {1992}}%
}]{%
Rees:1992me}
\APACinsertmetastar {%
Rees:1992me}%
\begin{APACrefauthors}%
Rees, M\BPBI J.%
\BCBT {}\ \BBA {} M{\'{e}}sz{\'{a}}ros, P.%
\end{APACrefauthors}%
\unskip\
\newblock
\APACrefYearMonthDay{1992}{sep}{},
\newblock
\unskip
\newblock
\APACjournalVolNumPages{Mon. Not. R. Astron. Soc.}{\bf 258}{1}{41P--43P}.
\newblock
\begin{APACrefDOI} \doi{10.1093/mnras/258.1.41P} \end{APACrefDOI}
\PrintBackRefs{\CurrentBib}

\bibitem [\protect \citeauthoryear {%
{Rueda-Becerril}%
}{%
{Rueda-Becerril}%
}{%
{\protect \APACyear {2020}}%
}]{%
RuedaBecerril:2020Paramo}
\APACinsertmetastar {%
RuedaBecerril:2020Paramo}%
\begin{APACrefauthors}%
{Rueda-Becerril}, J\BPBI M.%
\end{APACrefauthors}%
\unskip\
\newblock
\APACrefYearMonthDay{2020}{{\APACmonth{09}}}{},
\newblock
\APACrefbtitle {{Paramo: PArticle and RAdiation MOnitor}.} {{Paramo: PArticle
  and RAdiation MOnitor}.}
\PrintBackRefs{\CurrentBib}

\bibitem [\protect \citeauthoryear {%
J.~Rueda-Becerril%
}{%
J.~Rueda-Becerril%
}{%
{\protect \APACyear {2020}}%
}]{%
RuedaBecerril:2020kn}
\APACinsertmetastar {%
RuedaBecerril:2020kn}%
\begin{APACrefauthors}%
Rueda-Becerril, J.%
\end{APACrefauthors}%
\unskip\
\newblock
\APACrefYearMonthDay{2020}{}{},
\newblock
\APACrefnote{(In press)}
\PrintBackRefs{\CurrentBib}

\bibitem [\protect \citeauthoryear {%
J\BPBI M.~Rueda-Becerril%
}{%
J\BPBI M.~Rueda-Becerril%
}{%
{\protect \APACyear {2017}}%
}]{%
RuedaBecerril:2017phd}
\APACinsertmetastar {%
RuedaBecerril:2017phd}%
\begin{APACrefauthors}%
Rueda-Becerril, J\BPBI M.%
\end{APACrefauthors}%
\unskip\
\newblock
\APACrefYear{2017}.
\unskip\
\newblock
\APACrefbtitle {{Numerical treatment of radiation processes in the internal
  shocks of magnetized relativistic outflows}} {{Numerical treatment of
  radiation processes in the internal shocks of magnetized relativistic
  outflows}}\ \APACtypeAddressSchool {\BUPhD}{}{}.
\unskip\
\newblock
\APACaddressSchool {Valencia, Spain}{{Universitat de Val{\`e}ncia}}.
\PrintBackRefs{\CurrentBib}

\bibitem [\protect \citeauthoryear {%
{Rybicki}%
\ \BBA {} {Lightman}%
}{%
{Rybicki}%
\ \BBA {} {Lightman}%
}{%
{\protect \APACyear {1979}}%
}]{%
Rybicki:1979}
\APACinsertmetastar {%
Rybicki:1979}%
\begin{APACrefauthors}%
{Rybicki}, G\BPBI B.%
\BCBT {}\ \BBA {} {Lightman}, A\BPBI P.%
\end{APACrefauthors}%
\unskip\
\newblock
\APACrefYear{1979},
\newblock
\APACrefbtitle {Radiative processes in astrophysics} {Radiative processes in
  astrophysics}.
\newblock
\APACaddressPublisher{New York}{Wiley-Interscience}.
\PrintBackRefs{\CurrentBib}

\bibitem [\protect \citeauthoryear {%
Sari%
, Piran%
\BCBL {}\ \BBA {} Narayan%
}{%
Sari%
\ \protect \BOthers {.}}{%
{\protect \APACyear {1998}}%
}]{%
Sari:1998pi}
\APACinsertmetastar {%
Sari:1998pi}%
\begin{APACrefauthors}%
Sari, R.%
, Piran, T.%
\BCBL {}\ \BBA {} Narayan, R.%
\end{APACrefauthors}%
\unskip\
\newblock
\APACrefYearMonthDay{1998}{apr}{},
\newblock
\unskip
\newblock
\APACjournalVolNumPages{Astrophys. J.}{\bf 497}{1}{L17--L20}.
\newblock
\begin{APACrefDOI} \doi{10.1086/311269} \end{APACrefDOI}
\PrintBackRefs{\CurrentBib}

\bibitem [\protect \citeauthoryear {%
Woosley%
}{%
Woosley%
}{%
{\protect \APACyear {1993}}%
}]{%
Woosley:1993ga}
\APACinsertmetastar {%
Woosley:1993ga}%
\begin{APACrefauthors}%
Woosley, S\BPBI E.%
\end{APACrefauthors}%
\unskip\
\newblock
\APACrefYearMonthDay{1993}{mar}{},
\newblock
\unskip
\newblock
\APACjournalVolNumPages{Astrophys. J.}{\bf 405}{}{273}.
\newblock
\begin{APACrefDOI} \doi{10.1086/172359} \end{APACrefDOI}
\PrintBackRefs{\CurrentBib}

\bibitem [\protect \citeauthoryear {%
{Zhang}%
\ \BBA {} et al.%
}{%
{Zhang}%
\ \BBA {} et al.%
}{%
{\protect \APACyear {2020}}%
}]{%
Zhang:2020ch}
\APACinsertmetastar {%
Zhang:2020ch}%
\begin{APACrefauthors}%
{Zhang}, H.%
\BCBT {}\ \BBA {} et al.%
\end{APACrefauthors}%
\unskip\
\newblock
\APACrefYearMonthDay{2020}{{\APACmonth{06}}}{},
\newblock
\unskip
\newblock
\APACjournalVolNumPages{Mon. Not. R. Astron. Soc.}{\bf 496}{1}{974-986}.
\newblock
\begin{APACrefDOI} \doi{10.1093/mnras/staa1583} \end{APACrefDOI}
\PrintBackRefs{\CurrentBib}

\end{thebibliography}

\end{document}